\documentstyle{article}
\begin{document}
\LARGE
\begin{center}
\bf Real Tunneling and Black Hole Creation
\vspace*{0.6in}
\normalsize \large \rm 

Wu Zhong Chao

Dept. of Physics

Beijing Normal University

Beijing, 100875, China

Ref:\it Int. J. Modern Phys. \rm \bf D \rm 7, 111 (1998)

\vspace*{0.4in}
\large
\bf
Abstract
\end{center}
\vspace*{.1in}
\rm
\normalsize
\vspace*{0.1in}

We discuss the Hawking theory of quantum cosmology with regard to
approximation at the lowest order of  the Planck constant. 
At this level, the quantum scenario will be reduced to its 
classical evolutions in real and imaginary times. We restrict 
our attention to the so-called real tunneling case. It can be 
shown that, even at this level, there still 
exist some quantum effects, the classical field equation
may not hold at the transition surface. One can introduce the 
concept of constrained gravitational instanton. It may play 
some important role in the scenario of black hole creation in the
inflationary 
background at the Planckian era of the universe. From the
constrained gravitational instanton, the real 
tunneling can occur through different ways.  Consequently, it
will lead to the 
creation of different parts of the black hole spacetime in the
de Sitter background. The global aspects of the black hole
creation
are discussed.

\vspace*{0.3in}

PACS number(s): 98.80.Hw, 98.80.Bp, 04.60.Kz, 04.70.Dy

\vspace*{0.3in}

e-mail: wu@axp3g9.icra.it

\pagebreak

\large \bf I. Introduction
\vspace*{0.15in}

\rm 

\normalsize

The Hawking theory of  the No-Boundary Universe is the first
success in
obtaining a self-contained cosmology. Now,  in principle,
one can predict everything in the universe solely from the
theory.

According to the no-boundary proposal of Hartle and Hawking, the
quantum state of the universe is in the ground state, which is
defined by a path integral over compact Euclidean metrics [1]. 
Conceptually, this proposal is very appealing. Technically, like
any other theory of quantum cosmology, it 
encounters enormous difficulty in calculations due to the lack 
of a complete theory of quantum gravity. The task of this article
is to deal with the situation at the modest level, i.e, the 
lowest order in the Planck constant $\hbar$. It turns out that 
the problem is not really as trivial as it looks at the first 
glance.

In quantum field theory, it is well known that quantum tunneling
can be studied by using the instanton theory. Instanton is a sort
of
Euclidean solution of the field equation.  Quantum penetrating
can
be described by an analytic continuation from the Euclidean
solution
to its Lorentzian counterpart. In flat spacetime background, one
can
readily realize this by simply changing the time value from
imaginary to real.
However, except for some very special cases, a complex solution
of
the Einstein field equation does not typically have both purely
Euclidean and 
Lorentzian sectors. Therefore, the instanton theory 
cannot be used here without modification.

In some simple models, the creation of the universe can be
considered as a 
quantum tunneling from an Euclidean spacetime to a Lorentzian
one.  If  the
field equation is not only satisfied in the Euclidean and
Lorentzian spacetimes, 
respectively, but also at the location of the transition, then
the instanton theory
can be used as it is. Unfortunately, very few models in quantum 
cosmology share such luck [2]. The de Sitter model is an
exception.

The de Sitter model is the first nontrivial model in quantum
cosmology.
Since the 4-sphere and the de Sitter spacetime are the Euclidean
and Lorentzian sectors of a complex solution to the vacuum
Einstein
equation with a cosmological constant,  the instanton theory does
apply here. The main reason that the creation of the de Sitter 
universe can be considered as tunneling from the 4-sphere is that
it is
a model with only one degree of freedom, the scale of the
3-metric.
Indeed, the instanton theory can be used to all models with only
one degree of freedom.

However, if one discusses a model with more than one degrees of
freedom,
then the situation becomes more complicated. For more realistic 
models such as the Hawking massive scalar or primordial black
hole creation, the 
instanton theory has to be modified and generalized.

In the Hawking model a massive scalar
$\phi$ with mass $m$ is coupled to an isotropic and homogeneous 
universe [3]. It can be shown that there does not exist any
instanton solution
in this model. In fact, there does not exist any compact sector
of a complex
solution, as the 4-sphere solution to the de Sitter model.

Section II will review the Wheeler-DeWitt equation at the lowest
order. It is known that, in general,  the wave packet represents 
two ensembles of classical evolutions, one is in real time and
other in imaginary.
They satisfy classical equations with modification due to their
mutual interactions. These two ensembles of trajectories are
mutually orthogonal with respect to the supermetric of the 
configuration space.

Section III will deal with the modification of the instanton 
theory for the so-called real tunneling case.   That is, there is
no
interaction between Euclidean and Lorentzian evolutions, or the 
two ensembles decouple. To classically interpret the quantum 
tunneling, one has to begin with the action from first
principles.
Then we can introduce the concept of a constrained gravitational
instanton [4],
which is an Euclidean stationary action solution under some
constraints. It satisfies the Einstein equation with the possible
exception at the transition
surface where the constraints are imposed.

A very interesting application of the theory will be the
primordial 
black hole creation in quantum cosmology [5]. Although some
attempts were made one decade ago on the Schwarzschild-de Sitter
black hole creation in quantum cosmology [6][7], its conclusive
solution was obtained only very recently. This will be the
content 
of Section IV.  The problem of a black hole of the whole
Kerr-Newman
family in the de Sitter background has been completely resolved.
Its 
probability, at the $WKB$ level, is the exponential of a quarter 
of the sum of  the black hole and cosmological horizons. It turns
out that the de Sitter
evolution is the most probable trajectory in the Planckian era of
the
universe.
 
Section V shows that there are alternative ways of real tunneling
in the black hole 
creation with different probabilities. They will lead to the
Lorentzian evolutions of parts of interior of the black hole
horizon or 
exterior of the cosmological horizon. Section VI is devoted to
the global 
aspects of black hole creation. Section VII is a summary.

\vspace*{0.3in}

\large \bf II. The wave function at the lowest approximation

\vspace*{0.15in}

\rm 
\normalsize
In the No-Boundary Universe the wave function of the
universe is given by [1]
\begin{equation}
\Psi(h_{ij}, \phi) = \int_{C}d[g_{\mu \nu}] d [\phi] \exp (-
\bar{I} ([g_{\mu \nu}, \phi]),
\end{equation}
where the path integral is over the class $C$ of compact
Euclidean $4$-metrics and matter field configurations, which
agree with the given $3$-metrics $h_{ij}$ of the only boundary
and matter configuration $\phi$ on it. Here $\bar{I}$ means the
Euclidean action.

The Euclidean action for the gravitational part for a smooth
spacetime manifold $M$ with boundary $\partial M$ is
\begin{equation}
\bar{I} = - \frac{1}{16\pi} \int_M (R - 2\Lambda) -
\frac{1}{8\pi} \int_{\partial M} K,
\end{equation}
where $\Lambda$ is the cosmological constant, $R$ is the scalar
curvature and $K$ is the trace of the second fundamental form of
the
boundary.

The dominant contribution to the path integral comes from some
stationary action manifolds with matter fields on them, which are
the saddle points of the path integral. In general, the wave 
function takes a superposition form of wave packets
\begin{equation}
\Psi \approx C \exp (- S/\hbar),
\end{equation}
where we have written $\hbar$ explicitly; $C$ is a slowly varying
prefactor; and $S \equiv S_r + iS_i$ is a complex phase.

Since the wave packets of form (3) are not independent in the
decomposition of the wave function, one more restriction should
be imposed.  That is,  the wave packets itself should obey the
Wheeler-DeWitt equation. Classically, it means that the
evolutions represented by the wave packet should satisfy the
Einstein
equation with some quantum corrections, as it will be shown
below.

The Wheeler-DeWitt equation takes the following form module to
some operator ordering ambiguities:
\begin{equation}
\left(-\frac{1}{2} \bigtriangleup + V\right)\Psi = 0,
\end{equation}
where $\bigtriangleup$ is the Laplacian in the supermetric of the
configuration space and $V$ is considered as the potential term. 
Classically, it means that the evolutions represented by the wave
packet should satisfy the Einstein
equation with some quantum corrections. We have implicitly
assumed the universe to be closed.

Inserting the wave packet form into the Wheeler-DeWitt equation,
one obtains:
\begin{equation}
\left (- \frac{1}{2} (\bigtriangledown S)^2 + V \right )C +
\left( \frac{1}{2} 
\bigtriangleup S + \bigtriangledown S \cdot \bigtriangledown
\right)C\hbar - \frac{1}{2} \bigtriangleup C \hbar^2 = 0.
\end{equation}
We can separate it into a real part
\begin{equation}
\left (-\frac{1}{2}(\bigtriangledown S_r)^2 +
\frac{1}{2}(\bigtriangledown S_i)^2 + V\right )C +
\left(\frac{1}{2} 
\bigtriangleup S_r + \bigtriangledown S_r \cdot \bigtriangledown
\right) C\hbar - \frac{1}{2}\bigtriangleup C \hbar^2= 0,
\end{equation}
and an imaginary part
\begin{equation}
- C\bigtriangledown S_r \cdot \bigtriangledown S_i +
\left(\frac{1}{2}
C \bigtriangleup S_i + \bigtriangledown S_i \cdot
\bigtriangledown C\right)\hbar = 0.
\end{equation}

If we ignore the quantum effects represented by the terms
associated with powers of $\hbar$ in these equations, then Eqs.
(6) and (7) become
\begin{equation}
-\frac{1}{2} (\bigtriangledown S_r)^2 + \frac{1}{2} (
\bigtriangledown S_i)^2 + V = 0,
\end{equation}
and
\begin{equation}
\bigtriangledown S_r \cdot \bigtriangledown S_i = 0.
\end{equation}

Eq. (8) is the Lorentzian (or Euclidean) Hamilton-Jacobi
equation, with $S_i$ (or $S_r$) and $\bigtriangledown S_i$ (or
$\bigtriangledown S_r$) identified as the classical action and
the canonical momenta, respectively. One can define Lorentzian
(or
Euclidean) orbits along integral curves with
$\frac{\partial}{\partial t}\equiv \bigtriangledown S_i \cdot
\bigtriangledown$ (or $\frac{\partial}{\partial \tau}
\equiv \bigtriangledown S_r \cdot \bigtriangledown$). The wave
function represents an ensemble of classical trajectories. The
Lorentzian (or Euclidean) trajectories will trace out orbits in
the presence of the potential $V - \frac{1}{2} (\bigtriangledown
S_r)^2$ (or $-V - \frac{1}{2} (\bigtriangledown S_i)^2$).

Here, two kinds of quantum corrections are involved: one due to
the higher $\hbar$ terms of Eq. (6) and the other due to the
nonzero value of $\bigtriangledown S_r$ (or $\bigtriangledown
S_i$). For the Lorentzian case, if the modification is not
negligible, then the evolutions should deviate quite dramatically
from  classical dynamics. This deviation, which has
little effect on the short range behavior, may be crucial to the
global properties of the universe. At the lowest approximation
the 
first kind of corrections are neglected.

The quantity $\Psi^* \Psi$ or the factor $\exp (-2S_r)$ in the
wave packet can be interpreted as the relative probability of
the Lorentzian trajectories in the region of the configuration
space with varying $S_i$. From Eq. (9) we know that the
Lorentzian 
and Euclidean trajectories are mutually perpendicular, or $S_r$
remains constant along the
orbits at the lowest order in $\hbar$. The $\hbar$ term of Eq.
(7) represents the probability creation rates during the
Lorentzian evolutions, and the probabilities are
conserved if and only if the first term vanishes. The $\hbar$
term of Eq. (6) represents the dynamic effects of probability
creation due to the Euclidean evolution. Briefly speaking, the
evolution in real time is causal, while the evolution in
imaginary time is stochastic.

While imaginary time becomes a commonly accepted notion  in
quantum cosmology,
it has seemed not accepted yet by workers in other fields. Many
people on quantum optics are talking about light propagation
at a speed higher than $c$ when transpassing a classically
prohibited
region. They notice that the light takes zero real time lapse for
the 
tunneling. If one agrees that the time lapse is imaginary within
this region and the only observational effect of imaginary time
is
an exponential decay of the signal, then there would not be any
puzzle left [8].

Eqs. (8) (9) represents the Wheeler-DeWitt equation for quantum
cosmology at the lowest order approximation. The second order
partial functional differential equation has been degraded into
the first order one. One of the consequences of this
approximation 
is that the propagation property associated with the wavelike
equation of
the exact equation has been ignored at this level. We shall
restrict our following arguments
to the lowest order approximation unless otherwise stated.

In some models there exists a so-called Euclidean regime in
configuration space with a purely real phase in the wave packet.
At the boundary of this region a transition from Euclidean
evolution to Lorentzian evolution occurs through a 3-geometry
$\Sigma$. This transition is called real tunneling by Gibbons and
Hartle [2].

During real tunneling, the Euclidean spacetime is connected to
the
Lorentzian spacetime with common boundary $\Sigma$. Gibbons and
Hartle
argue that if the Einstein equation holds at the both sides of
$\Sigma$,
then the second fundamental form $K_{ij}$ has to
vanish from the both sides [2],
\begin{equation}
K_{ij} = 0.
\end{equation}
There is no such restriction on the normal derivative of the
matter field there. Since $\Sigma$ has a vanishing second
fundamental form, one can construct the vacuum instanton manifold
by joining the Euclidean manifold with its orientation reversal
across $\Sigma$.

In the next section we shall show that the presumption made by
Gibbons and Hartle at the transition surface is too restrictive,
it would
exclude many interesting phenomena in quantum cosmology,
in particular the scenario of black hole creation at the 
birth of the universe [5].

From the above argument, we learn that at the lowest level, for 
the real tunneling, the Lorentzian or Euclidean classical
equations 
are satisfied along the trajectories in the interiors of the
Euclidean
or Lorentzian regimes.

One may wonder whether at the lowest approximation the quantum
theory
can be fully reduced into the classical theory. We shall argue
that this
is not always the case. So we have to look closely at the
transition 
surface of the manifold, or the boundary of the two regimes in
the
configuration space. At some cases the classical equation will
not be recovered there, and consequently condition (10) may not
always 
hold there.

We shall deal with real tunneling cases in the rest of the paper
unless stated otherwise.

\vspace*{0.3in}

\large \bf III. The constrained gravitational instanton
\vspace*{0.15in}

\rm
\normalsize

The probability of the Lorentzian trajectory emanating from the
3-surface $\Sigma$ with the matter field $\phi$ on it can be
written as
\begin{equation}
P = \Psi^\star \Psi = \int_C d [g_{\mu \nu}] d[\phi] \exp(-
\bar{I} ([g_{\mu \nu}, \phi ]),
\end{equation}
where the class $C$ is composed by all no-boundary compact
Euclidean 4-metrics
and matter field configurations which agree with the given
3-metric $h_{ij}$ and
the matter field $\phi$ on $\Sigma$.

Here, we do not restrict the class $C$ to contain regular metrics
only, since the derivation from Eq. (1) to Eq. (11) has already
led to some jump discontinuities in the extrinsic curvature at
$\Sigma$. This point is crucial for the wave function of the
universe,
otherwise it would be impossible to factorize expression (11)
into the ground state definition (1).

The main contribution to the path integral in Eq. (11) is due to
the stationary action 4-metric, which meets all requirements on
the 3-surface $\Sigma$. At the $WKB$ level,
the exponential of the negative of the stationary action is the
probability of the corresponding Lorentzian trajectory.

From the above viewpoint, an extension of  the class $C$ to
include
metrics with some mild singularities is essential. Indeed, it is
recognized that,
 in some sense, the set of all regular metrics is not complete. 
For many cases, under the usual regularity conditions and the
requirements at the equator $\Sigma$, there may not exist any
stationary action metric, i.e. a gravitational instanton. It is
not
clear, how large the class $C$ should be. A necessary condition
for a
metric to be a member is that its scalar curvature should be
well-defined mathematically. It is reasonable to include metrics
with jump
discontinuities of extrinsic curvature and their degenerate
cases, that is, the conical or pancake singularities. For this
kind of singularity, the quantity $g^{1/2}R$ can be interpreted
as a
distribution-valued density [9].

If we lift the requirement on the 3-metric of the equator, then
the stationary action solution becomes the regular gravitational
instanton, as it satisfies the Einstein equation everywhere. Then
the 
Gibbons-Hartle condition (10) should hold at the equator. The
probability
of the corresponding trajectory takes stationary value; it may be
maximum, minimum or neither [10].

The wave packet represents an ensemble of Lorentzian
trajectories.
If the regular gravitational instanton has minimum action, then
the 3-metric from which the Lorentzian
evolution supposes to emanate is determined, and the most
probable 
trajectory is therefore singled out.  As a result, quantum
cosmology
fully realizes its prediction power: there is no degree of
freedom left, with the exception of physical time [10]. If one
does not use the 
instanton theory, the degree of freedom is reduced to half  by
the ground state proposal, roughly speaking, due to the
regularity
condition at the south pole of the Euclidean manifolds in the
path 
integral.

In general, the regularity conditions on the 4-metrics and the
requirements from the equator $\Sigma$ sometimes are so strong
that no gravitational instanton exists. The reason is that one
cannot
require the regularity condition and the given  3-metric at the
equator
simultaneously in the variational calculation.
Therefore, hopefully, one can only find a nonregular
gravitational instanton with some mild
singularities within the class $C$. If this is the case, then Eq.
(10) will no longer hold at the singularities. The probability
of the real tunneling will not be stationary when lifting the
3-metric requirements.

It has been proven [9] that a stationary action regular solution
keeps its status under the extension of  the class $C$. However,
if a stationary
action regular solution cannot be found, then it can probably be
expected with
some singularities at its equator among the class $C$.

One can rephrase this by saying that the solution obeys the
generalized Einstein
equation in the whole manifold. Since this result is derived from
first principles, and if one believes that Nature is quantum,
then
one should not feel upset about this situation.

\vspace*{0.3in}

\large \bf  IV. The creation of a black hole
\vspace*{0.15in}

\rm 
\normalsize

In this section, we shall apply the constrained gravitational 
instanton theory to the problem of primordial black hole. The
creation
of a black hole in the whole Kerr-Newman family has been resolved
[5].

In the Hawking model, the universe at the Euclidean and inflation
stage can be approximated by a $S^4$ space and the de Sitter
space with an effective cosmological constant 
$\Lambda  \equiv 3m^2 \phi^2_0$, where $\phi_0$ is the initial 
value of the scalar field. This is the motivation to 
discuss the black hole creation in the de Sitter background. 
A chargeless and nonrotating black hole sitting in the de Sitter
background can be described by the Schwarzschild-de Sitter
spacetime. It is the unique spherically symmetric vacuum solution
to the Einstein equation with a cosmological constant $\Lambda$.
The $S^2 \times S^2$ Nariai spacetime is its degenerate case.

Its Euclidean metric can be written as [11]
\begin{equation}
ds^2 = \left (1- \frac{2m}{r} - \frac{\Lambda r^2}{3} \right
)d\tau^2 
+\left (1- \frac{2m}{r} - \frac{\Lambda r^2}{3}\right )^{-1}dr^2 
+ r^2 d\Omega^2_2.
\end{equation}

For convenience one can factorize the potential [11]
\begin{equation}
\Delta = 1- \frac{2m}{r} - \frac{\Lambda r^2}{3}  = -
\frac{\Lambda}{3r}
(r - r_0)(r - r_2)(r - r_3).
\end{equation}

\begin{equation}
r_2 = 2\sqrt{\frac{1}{\Lambda}} \cos \left ( \alpha +
\frac{\pi}{3} \right ), \;\;\; r_3 = 2\sqrt{\frac{1}{\Lambda}}
\cos \left ( \alpha - \frac{\pi}{3} \right ),\;\;\;r_0 = -
2\sqrt{\frac{1}{\Lambda}} \cos \alpha ,
\end{equation}
and
\begin{equation}
\alpha = \frac{1}{3}\arccos (3m\Lambda^{1/2}),
\end{equation}
where $r_2, r_3$ are the black hole and cosmological horizons,
and $r_0$ is the horizon for the negative $r$. We are interested
in 
the Euclidean sector $r_2 \leq r \leq r_3$ for $0 \leq m \leq
 m_c = \Lambda^{-1/2}/3$. For the
extreme case $ m = m_c$ the sector 
degenerates into the $S^2 \times S^2$ space. 
 
The black hole and cosmological surface gravities $\kappa_2$ and
$\kappa_3$ are [11]
\begin{equation}
\kappa_2 = \frac{\Lambda}{6r_2}(r_3 - r_2)(r_2 - r_0),
\end{equation}
\begin{equation}
\kappa_3 = \frac{\Lambda}{6r_3}(r_3 - r_2)(r_3 - r_0).
\end{equation}

Now we are making a constrained gravitational instanton.
In the $(\tau - r)$ plane $r = r_2$ is an axis of symmetry,
the imaginary time coordinate $\tau$ is identified with period
$\beta_2 = 2\pi \kappa_2^{-1}$, and $\beta_2^{-1}$ is the
Hawking temperature. This makes the Euclidean manifold
regular at the black hole horizon. One can also apply this
procedure to the cosmological horizon with period $\beta_3 = 2\pi
\kappa_3^{-1}$, and $\beta^{-1}_3$ is the Gibbons-Hawking
temperature. For the $S^2 \times S^2$ case these
two horizons are identical, thus one obtains a regular
instanton. Except for the $S^2 \times S^2$ spacetime, one cannot
simultaneously regularize at both horizons. In fact, there is no
way to avoid singularity in compacting the Euclidean spacetime
because of the inequality $\beta_2^{-1} > \beta_3^{-1}$.

To form a constrained gravitational instanton [5], one can have
two
cuts at $\tau = consts.$ between $r = r_2$ and $r = r_3$ and then
glue them. Then
the $f_2$-fold cover turns
the $(\tau - r)$ plane into a cone with a deficit angle
$2\pi (1-f_2)$ at the black hole horizon. In a similar way one
can have an $f_3$-fold cover
at the cosmological horizon. Both $f_2$ and $f_3$ can take any
pair of real numbers with the relation 
\begin{equation}
f_2 \beta_2 = f_3 \beta_3
\end{equation}
for a fairly symmetric Euclidean manifold.

If $f_2$ or $f_3$ is different from $1$ (at least one should be),
then the cone at the black hole or cosmological horizon will have
an extra
contribution to the action of the manifold. We shall see that 
after the transition
to Lorentzian spacetime, the conical singularities will only
affect
the real part of the phase of the wave function, i.e. the
probability
of the black hole creation. The black hole creation can be
described by an analytic continuation  from imaginary time to 
real time of the constrained gravitational instanton  at the
equator which is two joint $\tau$ sectors, say $\tau = \pm f_2
\beta_2/4$ through the two horizons.

Since the integral of $K$ with respect to the $3$-area in the
boundary term of the action (2) is the area increase rate along
its normal, then the extra contribution due to the conical
singularities can be considered as the degenerate form shown
below
\begin{equation}
\bar{I}_{2,deficit} = - \frac{1}{8 \pi}\cdot 4\pi r_2^2\cdot 2\pi
(1 - f_2),
\end{equation}
\begin{equation}
\bar{I}_{3,deficit} = - \frac{1}{8 \pi}\cdot 4\pi r_3^2\cdot 2\pi
(1 - f_3).
\end{equation}

The volume term of the action for the manifold can be calculated
\begin{equation}
\bar{I}_{vol} = -\frac{\Lambda}{6} (r^3_3 - r^3_2) f_2 \beta_2.
\end{equation}

Using Eqs. (18) - (21), one can get the total action
\begin{equation}
\bar{I}_{total} = - \pi (r^2_2 + r^2_3).
\end{equation}
This is one quarter of  the negative of the sum of the two
horizon
areas. One quarter of the sum is the total entropy of the
universe.

It is remarkable to note that the action is independent of the
choice of $f_2$ or $f_3$. Our manifold satisfied the Einstein
equation
everywhere except for the two horizons at the equator. 
Consequently, the parameter
$f_2$ or $f_3$ is the only degree of freedom left.  In order to
check whether or not
 we get a stationary action solution or a constrained instanton,
one only needs to see
whether the above 
action is stationary with respect to this parameter. Our result
(22)
shows that our gravitational action has a stationary action, the
manifold is qualified as a constrained instanton,  and 
can be used for the $WKB$ approximation to the wave function. 
It also means no matter which flat fragment
of the constrained gravitational instanton is chosen, the same
black hole should be created with the same probability. Of
course, the most dramatic case is that of no volume, i.e. $f_2 =
f_3 = 0$.

Therefore, the probability of the black hole creation is 
\begin{equation}
P_m \approx \exp (\pi(r_2^2 + r_3^2)).
\end{equation}

Our result interposes  two special cases [12]. The first is the
de Sitter
model with $m = 0$,
\begin{equation}
P_0 \approx \exp \left ( \frac{3\pi}{\Lambda} \right )
\end{equation}
and the second is the Nariai model,  or pair black hole creation,
with
$m = m_c$,
\begin{equation}
P_c \approx \exp \left ( \frac{2\pi}{\Lambda} \right ).
\end{equation}

For the case $m \ll m_c$, we have
\begin{equation}
\bar{I}_{instanton} \approx - \pi \left [ \frac{3}{\Lambda} - 2m
\sqrt{\frac{3}{\Lambda}} + 2 m^2 \right ]
\end{equation}
and
\begin{equation}
P_m \approx P_0 \exp (- \pi r_3 r_2).
\end{equation}

For the case that $m$ is close to $m_c$, i.e. $\alpha
\approx 0$, one then has
\begin{equation}
\bar{I}_{instanton} \approx - \frac{2\pi}{\Lambda} (1 + 2
\alpha^2)
\end{equation}
and 
\[
P_m \approx \exp \left ( \frac{2\pi}{\Lambda}(1 +
2\alpha^2)\right ) 
\]
\begin{equation}
\approx P_{c} \exp \left ( \frac{4\pi \alpha^2}{\Lambda} \right
).
\end{equation}

The probability is an exponentially decreasing function in terms
of the mass parameter. The de Sitter case has the maximum 
probability and the Nariai case has the minimum probability.

The topology of the 3-metric of the equator is $S^2 \times S^1$,
the
configuration space has two degrees of freedom, one being  the
size of the universe (or the scale of $S^2$), the other being 
the mass parameter. 
This situation is very fortunate in  that the quantum creation of
a black hole can be realized by
a real tunneling. 

If one includes an electromagnetic field into the model, one
would 
be able to carry out a similar calculation. One simply replaces 
the potential by
\begin{equation}
\Delta = 1 - \frac{2m}{r} + \frac{Q^2}{r^2} - \frac{\Lambda
r^2}{3}
= - \frac{\Lambda}{3r^2}(r - r_0)(r - r_1)(r - r_2)(r - r_3),
\end{equation}
where $Q$ is the charge parameter of the black hole. 

For the magnetically charged black hole case, the configuration
of the
wave function is the 3-metric and magnetic charge. However,
the configuration for the wave function of an electrically
charged 
black hole is not well defined [5][13][14][15], if one naively
uses
the folding 
and gluing techniques described above. For the electric case the 
configuration of the wave function is the 3-metric and 
the canonical momentum conjugate to the charge. 
In order to get the wave function for the charge, one has to
appeal to a Fourier transformation, by which the duality between
electric and magnetic black holes is recovered.

The quantum creation scenario of a Schwarzschild-de Sitter black
hole
and the Reissner-Nordstr$\ddot{o}$m-de Sitter black hole
can be clearly depicted by using the so-called synchronous
coordinates [15]:
\begin{equation}
ds^2 = -d \eta^2 + \frac{1}{\cos^2 \xi } \left (
\left . \frac{\partial r}{\partial \xi } \right |_\eta
\right )^2 
d \xi^2 + r^2(\eta , \xi) (d \theta^2 + \sin^2
\theta d \phi^2).
\end{equation}
The Einstein constraint implies that
\begin{equation}
\dot{r}^2 + 1 - \frac{2m}{r} + \frac{Q^2}{r^2} - \frac{\Lambda
r^2}{3}
= \cos^2 \xi,
\end{equation}
where dot represents the derivative with respect to time, which 
is $\eta$ here, and $m$ is the integral constant, which is
identified
as the mass parameter.

The relations between coordinates $(t, r)$ and $(\eta, \xi )$
are:
\begin{equation}
\eta = \int \frac{dr}{(E^2 - \Delta )^{1/2}},
\end{equation}
\begin{equation}
t = \int \frac{Edr}{(E^2 - \Delta)^{1/2} \Delta},
\end{equation}
where
\begin{equation}
E = |\cos \xi |,
\end{equation}
and
\begin{equation}
\Delta = 1 - \frac{2m}{r} + \frac{Q^2}{r^2} -  \frac{\Lambda
r^2}{3}.
\end{equation}
\def\uptriangle{\Delta}

Its  classical evolution is 
equivalent to a coherent motion of particles along a congruence
of
timelike geodesics, labeled by $(\xi, \theta, \phi )$, in a
potential 
hill $\Delta $. One can release a particle from the potential
hill between
$r_2$ and $r_3$. Except for the case with  the initial position
at the top of the potential,
the particle will approach infinity or hit the singularity $r =
0$ for the chargeless black hole case. 
Therefore, the synchronous coordinates cover the whole spacetime
manifold. For the charged
case, the potential will blow up near $r = 0$, and the particle
will transpass the inner horizon
$r = r_1$ and is bounced back by the singularity at $r = 0$.

 By using this coordinate system one can obtain the 
wave function for the Schwarzschild-de Sitter and the
Reissner-Nordstr$\ddot{o}$m-de Sitter spacetimes [15] for the
spacelike 3-geometry
covered by the coordinates.

The whole scenario of the chargeless black hole creation is shown
in Fig. 1.
The $S^2$ space $(\theta - \phi )$ is represented by a $S^1$
space 
around the vertical axis. The radius of $S^2$ is $r$.
 The bottom part shows the instanton, the upper part shows the 
black hole created. The inner edge of the donut collapses from
the black horizon to the singularity $r=0$. The outer edge
expands from the cosmological horizon to $r = \infty$. The
conical singularities are not shown in
this 
very sketchy picture. The scenario of the charged black hole
creation is
similar except that the motion of the infalling trajectories are
bounced by the
singularity.

For the rotating and charged  black hole case, the spacetime
metric takes the Kerr-Newman form
\begin{equation}
ds^2 = \rho^2(\Delta^{-1}_r dr^2 + \Delta^{-1}_\theta d\theta^2)
+ \rho^{-2}
 \Xi^{-2}
\Delta_{\theta} \sin^2 \theta (adt - (r^2 + a^2) d\phi)^2 -
\rho^{-2} \Xi^{-2}\Delta_r  (dt - a \sin^2 \theta d \phi)^2,
\end{equation}
where
\begin{equation}
\rho^2 = r^2 + a^2 \cos^2 \theta,
\end{equation}
\begin{equation}
\Delta_r = (r^2 + a^2)(1 - \Lambda r^2 3^{-1}) - 2mr + Q^2 + P^2,
\end{equation}
\begin{equation}
\Delta_{\theta} = 1 + \Lambda a^2 3^{-1} \cos^2 \theta,
\end{equation}
\begin{equation}
\Xi = 1 + \Lambda a^2 3^{-1}
\end{equation}
and $m, a, Q$ and $P$ are constants, $m$ and $ma$ represent
mass and  angular momentum. $Q$ and $P$ are electric and
magnetic charges.

One can factorize $\Delta_r$ as follows:
\begin{equation}
\Delta_r = -\frac{\Lambda}{3} (r - r_0)(r - r_1)(r - r_2)(r -
r_3),
\end{equation}
where the roots $r_0, r_1, r_2$ and $r_3$ are in ascending order,
$r_2$ and
$r_3$ are the black hole and cosmological horizons. 

To form the constrained gravitational instanton, one can
imaginarize 
the time coordinate by setting $ t = -i\tau$, and then do folding
and 
gluing as in the nonrotating cases.

If one were to naively factorize the probability from Eq. (11),
he would
get the wave function for the 3-metric and the differential
rotation of two horizons only. So one has 
to use another Fourier transformation to obtain the wave function
for angular momentum [5]. If  the hole is electrically
charged, one has to appeal again to the Fourier transformation,
as we did for the
nonrotating case. Here, two Fourier transformations are involved.

At any case, the probability of a black creation in the de Sitter
background, at the $WKB$ level, is the exponential of a quarter
of the sum of  the black hole and cosmological horizons, a
quarter
of the sum is the total entropy of the universe. By the no-hair 
theorem, the problem of a single black hole creation in quantum
cosmology has been resolved completely.

The probability is an exponentially decreasing function of mass,
charge magnitude and angular momentum. The de Sitter evolution
is the most probable one at the Planckian era [5].

\vspace*{0.3in}

\large \bf V. The alternative tunnelings

\vspace*{0.15in}

\rm 
\normalsize

Now we are going to discuss a black hole creation from an
alternative
route. We begin with the vacuum Kantowski-Sachs model [16] with
the 
positive cosmological constant. The 3-surface is homogeneous and 
has topology $S^1 \times S^2$. The treatment in this section is
suitable 
for the Kaluza-Klein $S^1 \times S^n$ case [10], but we shall
study 
the $n = 2$ case only below for simplicity. In Ref. [16] a black
hole
will be formed when the massive scalar field rolls down the
potential hill and 
starts to oscillate. The effect of the massive scalar field is
approximated by a cosmological constant
during the imaginary time stage and the inflationary stage. 
Here, we investigate the case of a black hole
creation at the exact moment of the birth of the universe.

The Euclidean metric of the Kantowski-Sachs model takes the form:
\begin{equation}
ds^2 = d\tau^2 + a^2( \tau ) d\omega^2 + b^2( \tau ) d
\Omega^2_2,
\end{equation}
where $\omega$ is identified with a period $2\pi$.

The Euclidean field equation is
\begin{equation}
b  \ddot{b} + \frac{{\dot{b}}^2}{2} - \frac{1}{2}
+ \frac{\Lambda b^2}{2} = 0,
\end{equation}
\begin{equation}
\ddot{a} b + \ddot{b} a + \dot{a} \dot{b} 
+ \Lambda ab = 0
\end{equation}
and
\begin{equation}
b \dot{b} \dot{a} + \frac{a \dot{b}^2}{2} - \frac{a}{2}
+ \frac{\Lambda ab^2}{2}= 0,
\end{equation}
where the dot means the derivative with respect to $\tau$.

There are two kinds of gravitational instantons. They correspond
to the
regularity condition at the south pole. (I): $a = a_0, \dot{a} =
0,
b = 0 $ and $ \dot{b} = 1$; (ii):  
$b = b_0, \dot{b} = 0, a = 0
$ and $ \dot{a} = 1$. 

The gravitational instanton with the boundary condition (i) is
\begin{equation}
ds^2 = d\tau^2 + a_0^2 \cos^2 H\tau d\omega^2 + H^{-2} 
\sin^2 H\tau d \Omega^2_2,
\end{equation}
where $H = \sqrt{\Lambda / 3}$ is the Hubble constant.

After the scale of $S^2$ reaches maximum, then one can make an
analytical continuation along real time direction to get its 
Lorentzian counterpart
\begin{equation}
ds^2 = -dt^2 + a_0^2 \sinh^2 Ht d\omega^2 + H^{-2}\cosh^2 Ht
d\Omega^2_2.
\end{equation}

The Lorentzian metric is a part of the de Sitter
space with a cone singularity at the birth of the universe due to
the identification of the circle $S^1$ with the finite period.
The Einstein
equation does not hold at the transition as $\tau = \pi/2H$ and
$t = 0$
unless at the Euclidean side as $ a_0 = H^{-1}$. The 
solution has a scale invariance associated with the circle. Thus 
there essentially exists only one trajectory. This mild
singularity is
acceptable. The fact that there are so many beautiful
representations 
for the de Sitter spacetime is a manifestation of its
versatility.

The gravitational instanton with the boundary condition (ii)  is
the Schwarzschild-de Sitter manifold of the last section, in
which $r$ is 
identified as $b$, the size of $S^2$, $\Delta $ is identified
as $a^2$, and $\Delta^{-1}dr^2$ becomes $d\tau^2$ here.

\def\gt{>}

If one regularizes the black hole horizon by setting $f_2 = 1$ 
as in the last section, then one can consider the horizon as the
south
pole. 
It is noted that the expansion rate of $S^2$ space vanishes at
both
horizons. The spacetime becomes Lorentzian as one enters the
exterior of the cosmological horizon $r \gt r_3$.  Here, the $r$
coordinate
becomes timelike. At the transition there is a conical
singularity,
while the scale of $S^1$ shrinks to zero. The Lorentzian
evolution represents a part of the exterior of cosmological
horizon
with $S^1$ of extension $\Delta t = \beta_2$. The constrained
gravitational instanton is formed by joining the above manifold
with its
orientation reversal across the cosmological horizon. Therefore,
the 
action should be twice as much as the action of the constrained
instanton discussed in the last section. The probability of the
creation,
at the $WKB$ level, is
\begin{equation}
P_3 \approx \exp (2\pi (r^2_2 + r^2_3)).
\end{equation}

On the other hand, if one lets $f_3 = 1$, then the cosmological
horizon
can be considered as the south pole  and quantum transition will
occur at
the black horizon. The Lorentzian evolution describes an interior
part
of the black hole. One just exchanges the two horizons in the
statement of the preceding paragraph.
One will obtain the same creation probability at the $WKB$ level
\begin{equation}
P_2 \approx \exp (2\pi (r^2_2 + r^2_3)).
\end{equation}

The manifolds created by the alternative tunnelings are shown as
regions
$OGH$ and $EMN$ in Fig. 2, respectively. The curves $OG$ and $OH$
(curves $EM$ and $EN$) are identified due to the periodicity
condition
required by Euclidean compactification at the cosmological (black
hole) 
horizon.

To apply the above arguments on the alternative tunnelings to the
black holes of the whole
Kerr-Newman family is straightforward.

\vspace*{0.3 in}

\large \bf VI.  Global aspects of the black hole creation

\normalsize

\vspace*{0.3in}

\rm

Now we discuss the global property of the black hole creation.

The synchronous coordinates cover the whole manifold of  the 
Schwarzschild-de Sitter spacetime. This can be shown clearly by
the 
Penrose-Carter diagrams [11] in Fig.2. There is an infinite
sequence 
of diamond shape regions, singularities 
$r = 0$ and spacelike infinities $r = \infty$. Therefore, the
3-geometry
is not closed and not suitable for No-Boundary Universe. However,
in the scenario of  a black hole creation
the whole manifold created can be obtained by an identification,
for
instance, with lines $ABC$ and $DEF$. The equator in the
instanton is identified
as the closed line $BE$ here. 

It is worth mentioning that $t$ coordinate is future (past)
directed 
in the right (left) triangle above line $BE$. This is consistent
with the analytic continuation of the imaginary time at the two
cuts
$\tau = consts$ for the south hemisphere of the constrained
gravitational
instanton, or the two ends of the imaginary time lapse. The 
same argument applies to the other members of the Kerr-Newman
family. 
In the Schwarzschild space, $r_3$ becomes
$r = + \infty$, and no identification is necessary.

Fig. 3 shows the Penrose-Carter diagram for 
Reissner-Nordstr$\ddot{o}$m-de Sitter black hole creation. 
The lines $ABC$ and $DEF$ are identified. The
line $BE$ is the quantum transition surface, from which the
region
of $HABOEDLG$ is created. The synchronous coordinates cover the
whole manifold
minus  neighborhoods of the singularity.  One may travel to other
universes 
by passing through the `` wormholes" made by the charge. In
particular, if one follows
an infalling trajectory in the synchronous coordinates, he will
be bounced by
the singularity and enter the another universe. But this does not
bother us right now, since
the relevant 3-metric will no longer be spacelike, and the wave
function is not
well defined under this circumstance.  In the
Reissner-Nordstr$\ddot{o}$m space, $r_3$ becomes
$r = + \infty$, and no identification is necessary. 

Fig. 4 shows the Penrose-Carter diagram of the symmetry axis of
the Kerr-Newman-de Sitter black hole creation. The infinities
$r = + \infty $ and $r = - \infty$ are not joined together.
The open circles mark where the ring singularity occurs,
although it is not on the symmetry axis. 
The lines $ABC$ and $DEF$ are identified. The
line $BE$ is the quantum transition surface, from which the
region
of $HABOEDLG$ is created. In the Kerr-Newman space, $r_3$ becomes
$r = + \infty$, no identification is necessary.

The alternative tunnelings are shown by the shaded bands in Figs.
2 and 3
with the two boundaries identified. It should be similar for the
Kerr-Newman case.

\vspace*{0.3 in}

\large \bf VII. Summary

\normalsize

\vspace*{0.3in}

\rm

The complex tunnelings are common phenomena in Nature. We
investigate the
real tunneling problem in this paper. However the conventional
concept
of real tunneling is too narrow. If one insists on this, then
many
interesting phenomena, such as creation of a single black hole in
quantum cosmology, would be
excluded from studying,. On the other hand, if 
one begins with the first principles in quantum framework, it is
naturally leads to the constrained
gravitational instanton. The field equation holds in the
instanton except
for the location where the constraint is imposed; here is where
one will find the equator of the
instanton. If one believes that the nature is quantum, then the
field
equation becomes secondary, and one should welcome this kind of
generalization.

As an example, we have shown that the real tunnelings occur at
the quantum
creation of a black hole in the de Sitter background. This is a
quite rare
case in nature, taking into consideration  the fact that the
model has more than one degree of
freedom.

By the analysis of this article, one learns that real tunnelings
of quantum transition originates from the same 
Schwarzschild-de Sitter instanton in several ways. No matter what
kind
analytic continuation is made, one always obtain the whole or
parts of the 
same Lorentzian spacetime, but with different probabilities. 
If we are working with the de Sitter model,
the situation is not so transparent, since it is of the maximum
symmetry. All equators of the  4-sphere are identical. 

It is also interesting to examine how the Euclidean manifold is
joined to the
Lorentzian counterpart in the black hole creation. The Lorentzian
spacetimes have to be compactified by a periodic identification,
then the
global aspects of the whole scenario is clarified.

\vspace*{0.1in}

\bf Acknowledgment:

\vspace*{0.1in}
\rm

I would like to thank S.W. Hawking of Cambridge University and K.
Sato of
Tokyo University for their hospitality. I am grateful to G.W.
Gibbons for discussions.
 
\vspace*{0.1in}

\bf References:

\vspace*{0.1in}
\rm

1. J.B. Hartle and S.W. Hawking, \it Phys. Rev. \rm \bf D\rm
\underline{28}, 2960 (1983).

2. G.W. Gibbons and J.B. Hartle, \it Phys. Rev. \rm \bf D\rm
\underline{42}, 2458 (1990).

3. S.W. Hawking, \it Nucl. Phys. \rm \bf B\rm \underline{239},
257 (1984).

4. In the earlier version of this work, I used the term
`` generalized
gravitational instanton". 

5. Z.C. Wu, \it Int. J. Mod. Phys. \rm \bf D\rm\underline{6}, 199
(1997).

6. L.Z. Fang and M. Li, \it Phys. Lett. \rm \bf B\rm
\underline{169}, 28 (1986).

7. Z.C. Wu, in \it Fourth Marcel Grossman Meeting, \rm ed. R.
Ruffini (North Holland, Amsterdam, 1986).

8. Z.C. Wu, unpublished.

9. G. Hayward and J. Louko, \it Phys. Rev. \rm \bf D\rm
\underline{42}, 4032 (1990).

10. X.M. Hu and Z.C. Wu, \it Phys. Lett. \rm \bf B\rm
\underline{149}, 87 (1984).

11. G.W. Gibbons and S.W. Hawking, \it Phys. Rev. \bf D\rm
\underline{15}, 2738 (1977).

12. R. Bousso and S.W. Hawking, \it Phys. Rev. \rm \bf D\rm
\underline{52}, 5659 (1995).

13. R.B. Mann and S.F. Ross, \it Phys. Rev. \rm \bf D\rm
\underline{52}, 2254 (1995).

14. S.F. Ross and S.W. Hawking, \it Phys. Rev. \rm \bf D\rm
\underline{52}, 5865 (1995).

15. Z.C. Wu,  \it Prog. Theo. Phys. \rm 
\underline{97}, 859 (1997);  \it Prog. Theo. Phys. \rm 
\underline{97}, 873 (1997).

16. R. Laflamme and E.P.S. Shellard, \it Phys. Rev. \rm \bf D\rm
\underline{35}, 2315 (1987).
 
\newpage

Figure captions:

Fig. 1: The whole scenario of the black hole creation.
The $S^2$ space $(\theta - \phi )$ is represented by a $S^1$
space 
around the vertical axis. The radius of $S^2$ is $r$.
 The bottom part shows the instanton; the upper part shows the 
black hole created.
 
Fig. 2: The Penrose-Carter diagram for a Schwarzschld-de Sitter
black hole
creation.

Fig. 3: The Penrose-Carter diagram for a
Reissner-Nordstr$\ddot{o}$m-de
Sitter black hole creation.

Fig. 4: The Penrose-Carter diagram of the symmetry axis of
the Kerr-Newman-de Sitter black hole creation.
\vspace*{0.3 in}

\end{document}